\documentclass[submission,copyright,creativecommons]{eptcs}
\providecommand{\event}{TARK 2021} 
\usepackage{breakurl}             
\usepackage{underscore}           

\usepackage{latexsym}
\usepackage{amssymb, amsmath}

\title{Reasoning about Emergence of Collective Memory}
\author{R. Ramanujam
\institute{Institute of Mathematical Sciences}
\institute{Homi Bhabha National Institute \\
Chennai, India}
\email{jam@imsc.res.in}
}
\def\titlerunning{Emergence of Collective Memory}
\def\authorrunning{R. Ramanujam}
\begin{document}
\maketitle

\newcommand{\tree}{\ensuremath{\mathcal{T}}}
\newcommand{\treeset}{\ensuremath{\tau}}
\newcommand{\arena}{\ensuremath{\mathcal{A}}}
\newcommand{\game}{\ensuremath{\mathcal{G}}}
\newcommand{\acttuple}{\ensuremath{\mathbf{A}}}
\newcommand{\atuple}{\ensuremath{\mathbf{a}}}
\newcommand{\plays}{\ensuremath{P}}
\newcommand{\hists}{\ensuremath{H}}
\newcommand{\types}{\ensuremath{T}}
\newcommand{\play}{\ensuremath{\rho}}
\newcommand{\hist}{\ensuremath{\rho}}
\newcommand{\move}[1]{\ensuremath{\stackrel{#1}{\rightarrow}}}
\newcommand{\last}{\ensuremath{\mathit{last}}}
\newcommand{\occ}{\ensuremath{\mathit{occ}}}
\newcommand{\infi}{\ensuremath{\mathit{inf}}}
\newcommand{\prop}{\ensuremath{\mathcal{P}}}
\newcommand{\val}{\ensuremath{\mathit{val}}}
\newcommand{\win}{\ensuremath{\phi}}
\newcommand{\colour}{\ensuremath{\chi}}
\newcommand{\muller}{\ensuremath{\mathcal{F}}}
\newcommand{\reward}{\ensuremath{r}}
\newcommand{\strat}{\ensuremath{s}}
\newcommand{\stratset}{\ensuremath{\Sigma}}
\newcommand{\payoff}{\ensuremath{u}}
\newcommand{\util}{\ensuremath{u}}
\newcommand{\pref}{\ensuremath{\sqsubseteq}}
\newcommand{\stratuple}{\ensuremath{\mathbf{s}}}
\newcommand{\fst}{\ensuremath{\mathcal{Q}}}
\newcommand{\run}{\ensuremath{r}}
\newcommand{\lang}{\ensuremath{\mathcal{L}}}
\renewcommand{\path}{\ensuremath{\rho}}
\newcommand{\proj}{\ensuremath{\pi}}
\newcommand{\Obs}{\ensuremath{\Phi}}
\newcommand{\obs}{\ensuremath{\varphi}}
\newcommand{\diamondmin}{\Diamond\kern-0.54em{\raisebox{.25ex}{\rm -}}\kern0.175em}
\newcommand{\boxmin}{\Box\kern-0.54em{\raisebox{.25ex}{\rm -}}\kern0.175em}
\newcommand{\minus}[1]{\ensuremath{\langle #1 \rangle^-}}
\newcommand{\plus}[1]{\ensuremath{\langle #1 \rangle^+}}
\newcommand{\cl}{\ensuremath{\mathit{CL}}}
\newcommand{\at}{\ensuremath{\mathit{AT}}}
\newcommand{\md}{\ensuremath{\mathbf{\mathit{md}}}}
\newcommand{\mc}{\ensuremath{\mathcal{M}}}
\newcommand{\tp}{\ensuremath{\mathcal{TP}}}
\newcommand{\pt}{\ensuremath{\mathcal{PT}}}
\newcommand{\chop}{\ensuremath{\frown}}
\newcommand{\sembrack}[1]{[\![#1]\!]}
\newcommand{\qed}{\ensuremath{\Box}}
\newcommand{\sub}{\ensuremath{\mathit{sub}}}
\newcommand{\subf}{\ensuremath{\mathit{sf}}}
\newcommand{\sarena}{\ensuremath{\mathcal{Z}}}
\newcommand{\rest}{\ensuremath{\!\upharpoonright\!}}
\newcommand{\id}{\ensuremath{\mathit{id}}}
\newcommand{\io}{\ensuremath{O}}
\newcommand{\bigo}{\ensuremath{\mathcal{O}}}
\newcommand{\prob}{\ensuremath{\mathit{prob}}}
\newcommand{\tpl}[1]{\ensuremath{\langle #1 \rangle}}
\newcommand{\nbd}{\ensuremath{\mathit{nbd}}}
\newcommand{\nbdfam}{\ensuremath{\mathcal{X}}}
\newcommand{\pot}{\ensuremath{\phi}}
\newcommand{\clq}{\ensuremath{\mathit{clq}}}
\newcommand{\unf}{\ensuremath{\mathcal{C}}}
\newcommand{\precon}{\ensuremath{\mathit{pre}}}
\newcommand{\rimplies}{\ensuremath{\supset}}
\newcommand{\Rules}{\ensuremath{R}}
\newcommand{\subarenas}{\ensuremath{SA}}
\newcommand{\homo}{\ensuremath{h}}
\newcommand{\bool}{\ensuremath{\mathit{bool}}}
\newcommand{\true}{\ensuremath{\top}}
\newcommand{\false}{\ensuremath{\bot}}
\newcommand{\test}{\ensuremath{\psi}}
\newcommand{\homospecs}[2]{\ensuremath{\homo_{#1:#2}}}
\newcommand{\restauto}{\ensuremath{\mathcal{Q}}}
\newcommand{\stratauto}{\ensuremath{\mathcal{S}}}
\newcommand{\master}{\ensuremath{\mathcal{M}}}
\newcommand{\turn}{\ensuremath{\mathit{turn}}}
\newcommand{\proset}{\ensuremath{\mathbf{Y}}}
\newcommand{\pro}{\ensuremath{\mathbf{y}}}
\newcommand{\perm}{\ensuremath{\pi}}
\newcommand{\soc}{\ensuremath{\mathit{soc}}}
\newcommand{\plr}{\ensuremath{\mathit{plr}}}
\newcommand{\rational}{\ensuremath{\mathbb{Q}}}
\newcommand{\atset}{\ensuremath{\mathscr{C}}}
\newcommand{\mixed}{\ensuremath{\mathbf{x}}}
\newcommand{\mixedset}{\ensuremath{\mathbf{X}}}
\newcommand{\R}{\ensuremath{\mathcal{R}}}
\newcommand{\step}[1]{\ensuremath{\stackrel{#1}{\rightarrow}}}
\newcommand{\strata}{\ensuremath{\strat_0}}
\newcommand{\stratb}{\ensuremath{\strat_1}}
\newcommand{\Strata}{\ensuremath{\stratset_0}}
\newcommand{\Stratb}{\ensuremath{\stratset_1}}
\newcommand{\cert}{\ensuremath{\beta}}
\newcommand{\memupdate}{\ensuremath{\delta}}
\newcommand{\actupdate}{\ensuremath{g}}

\newcommand{\andover}{\displaystyle \bigwedge}
\newcommand{\orover}{\displaystyle \bigvee}
\newcommand{\capover}{\displaystyle \bigcap}
\newcommand{\cupover}{\displaystyle \bigcup}

\newcommand{\nat}{{\bf N}}

\newcommand{\mycom}[1]{\marginpar{#1}}
\newcommand{\GG}{\ensuremath{\mathcal{G}}}
\newcommand{\struct}{\ensuremath{\mathfrak{A}}}
\newcommand{\VV}{\ensuremath{\mathcal{V}}}
\newcommand{\vis}{\ensuremath{\mathit{vis}}}
\newcommand{\term}{\ensuremath{\mathbf{\tau}}}
\newcommand{\sees}{\ensuremath{\leftrightarrow}}
\newcommand{\num}{\ensuremath{\sharp}}
\newcommand{\varmap}{\ensuremath{\eta}}
\newcommand{\init}{\ensuremath{\mathit{init}}}
\newcommand{\reach}{\ensuremath{\mathit{reach}}}
\newcommand{\tm}{\ensuremath{\mathit{typ}}}
\newcommand{\out}{\ensuremath{\mathit{out}}}
\renewcommand{\types}{\ensuremath{\Gamma}}
\newcommand{\type}{\ensuremath{\gamma}}
\newcommand{\agraph}{\ensuremath{\mathcal{A}}}
\newcommand{\fin}{\ensuremath{\mathit{fin}}}
\newcommand{\clear}{\ensuremath{\mathit{clear}}}
\newcommand{\trest}{\ensuremath{\widetilde\tree}}
\newcommand{\crit}{\ensuremath{\mathit{crit}}}
\newcommand{\cls}{\ensuremath{\mathit{cls}}}

\newenvironment{replemma}[1]{\noindent {\bf Lemma~#1~}}{\smallskip}
\newenvironment{reptheorem}[1]{\noindent {\bf Theorem~#1~}}{\smallskip}
\newenvironment{repcor}[1]{\noindent {\bf Corollary~#1~}}{\smallskip}
\newenvironment{repclm}[1]{\noindent {\bf Claim~#1~}}{\smallskip}
\newenvironment{reppro}[1]{\noindent {\bf Proposition~#1~}}{\smallskip}


\newtheorem{theorem}{Theorem}
\newtheorem{proposition}{Proposition}
\newtheorem{lemma}{Lemma}
\newtheorem{claim}{Claim}
\newtheorem{definition}{Definition}
\newtheorem{cor}{Corollary}
\newtheorem{ass}{Assumption}
\newtheorem{question}{Question}

\newenvironment{proof}{\vspace{1ex}\noindent{\bf Proof}\hspace{0.5em}} 
	{\hfill$\qed$\vspace{1ex}}
\newenvironment{pfoutline}{\vspace{1ex}\noindent{\bf Proof
outline}\hspace{0.5em}} 
	{\hfill$\qed$\vspace{1ex}}
\newenvironment{proofof}[1]{\vspace{1ex}\noindent{\bf Proof of #1}\hspace{0.5em}} 
	{\hfill$\qed$\vspace{1ex}}
\newenvironment{rem}[1]{\vspace{1ex}\noindent{\bf Remark}\hspace{0.5em}}{ }

%

%
\begin{abstract}
We offer a very simple model of how collective memory may form.
Agents keep signalling within neighbourhoods, and depending on
how many support each signal, some signals ``win'' in that
neighbourhood. By agents interacting between different
neighbourhoods, `influence' spreads and sometimes, a collective
signal emerges. We propose a logic in which we can reason about
such emergence of memory and present preliminary technical results
on the logic.
\end{abstract}

\section{Introduction}
\begin{quote}
{\em Strictly speaking, there is no such thing as collective memory -- part of the 
same family of spurious notions as collective guilt. But there is collective 
instruction $\ldots$ All memory is individual, unreproducible; it dies with 
each person. What is called collective memory is not a remembering 
but a {\em stipulating}: that this is important, and this is the story about 
how it happened, with the pictures that lock the story in our minds.
}

\hfill {\sf Susan Sontag} (\cite{Son})
\end{quote}

Any discussion on individual values and social values visits the question,
{\em How are {\bf we} to act?} at some point. The question, of course, is,
who is this {\bf we} referred to here ? Clearly this {\bf we} is a social 
construction, one that depends on the very social norms and social values 
that we wish to reason about (\cite{Tuo}). Integral to such social construction of a
collective is the {\sf memory} ascribed to that collective. Group identity
is constructed structurally by ascribing memory to the group, and in turn,
such identity shapes its memory. Remembrance has a crucial impact on 
preferences and values, influences action.

It is here that Susan Sontag's quote above assumes significance. Sontag
calls collective memory a process of {\sf stipulation}. Somehow the collective
ascribes importance to an item of memory, authenticates it and symbolizes
it; then on, the symbolism ``locks'' the memory item, in Sontag's account.

Note that this is a significant departure from the structural conception 
of memory, that visualises memory as a notebook, and remembering as looking
it up.  Wittgenstein (\cite{Witt}) strongly attacked such a conception of memory. 

\begin{quote}

I saw this man years ago: now I have seen him again, I recognize him,
I remember his name. And why does there have to be a cause of this
remembering in my nervous system? Why must something or other, whatever
it may be, be stored up there in any form? Why must a trace have been
left behind? Why should there not be a psychological regularity to which
no physiological regularity corresponds?

If this upsets our concept of causality then it is high time it was upset.

\end{quote}

Scholars like Sutton (\cite{Sutt}) have discussed this issue at length.
For Wittgenstein, social acts were important in shaping memory, and based
on this, scholars like Rusu  (\cite{Rusu}) even talk of {\em social time},
and modern theories of connectionism and distributed memory build on 
many such notions. 

For us, these remarks are relevant from two viewpoints. The 1950's saw the 
development of automata theory as a study of {\em memory structures}, and in
theory of computation, automata provide a model of memory that Wittgenstein 
might have approved of. In this view, memory is not a table to be looked up,
but is constituted by states of being of the automaton. Observations cause
changes in state, some states remember (some of the past) and some forget.
Thus, remembering and forgetting are built into system structure. Such a view is important for
seeing memory and reasoning as {\em interdependent} rather than as separate
(as psychologists used to consider). Logicians are used to equating automata
and logics, as in the case of monadic second order logics of order or in the
case of Pressburger arithmetic. (Wittgenstein would have approved.) 

The other viewpoint relates to {\em distributed memory}, where interacting agents
rely on memory external to them. Computer science has evolved impressive
models of highly flexible interaction and memory that has literally changed
the everyday life of much of humanity in the last few decades. Today, memory
storage on the ``cloud'' has become indispensable for many, and people
voluntarily `post' personal information to make it socially available in an
attempt to write personal information into social memory. 

In social theory, the notion of collective memory is influential.  {\em Maurice 
Halbwachs} (\cite{Halb}) talked of how an individual's understanding of the past
is strongly linked to a group consciousness, which in turn is a form
of {\em group memory} that lives beyond the memories of individuals that
form the group. 

For the logician, these notions pose an interesting challenge: what are the
{\sf logical} properties of collective remembering? What is the rationale 
followed by a group in ascribing / stipulating collective importance to
events and their remembering? Why is a particular idealisation chosen?
These are difficult questions to answer, but a more modest reformulation 
of such questions offers an approach to solutions.  If the memory of an 
automaton is describable via logic, we can perhaps build a model of group 
and individual memory based on automata whose interactions lead to 
collective memory, which in turn influences behaviour of individual automata. 

Why should one bother? In (\cite{Pa}) Rohit Parikh speaks of {\em cultural
structures} providing an infrastructure to social algorithms (much as data
structures do for computational algorithms). Epistemic reasoning is an 
essential component of social algorithms, as persuasively argued by 
Parikh. We can then see collective memory as an essential gradient 
of its infrastructure creating the `common ground' in which social 
objectives and communications are interpreted (\cite{Clark}). Moreover, social 
algorithms such as elections need social memory if they are to
achieve their democratic purpose.

Moreover, there has been extensive research in recent years on notions
of  collective agency (\cite{Tamm}), collective action (\cite{RP-G}),
collective belief (\cite{Gilb}) and many more. However, the memory
required for collective action, belief and agency is largely assumed
rather than explicitly discussed. While social theorists extensively
discuss the role of society's needs for remembrance (as a way of
acknowledging the past and taking responsibility for it) the notion
of memory as an infrastructural need for such functions is often
glossed over. Moreover, the notions of memory that inform these
discussions are embodied in text, icons and physical tokens, much
like the notebook that Wittgenstein refers to (and objects to).
Contemporary reality, with the extension of human memory using
technology, suggests that more dynamic, behavioural models of
memory, and the re-inforcing behaviour that leads to stipulation
as referred to by Sontag, may be relevant.

What follows is a very simple, perhaps very simplistic, attempt at formalization
of this notion, inspired by the study of {\em population protocols} in distributed 
computing (\cite{Pop}) and large anonymous games \cite{DP}. We offer this 
formal model tentatively, as an initial step of a (hopefully) detailed research 
programme. As it stands, the model has no agency or epistemic attitudes or 
social aggregation. Instead it focusses only on how local attempts at signalling 
``importance'' of an observation can spread and lead to some stable 
phenomenon that can be meaningfully construed as collective memory.

The crucial element here is that individuals perceive events differently, based
on social and cultural background. The word {\em partition} would evoke 
the image of equivalence classes to a combinatorist in general, but very likely, 
that  of a terrible tragedy first to an Indian combinatorist. 

\subsection{Related work}
The literature on memory studies principally consists of two strands,
one on individual memory, substantially incorporated into psychology, and
the other on group memory, primarily on social representations of history
(\cite{Barash}. A major question of interest is whether such groups manifest 
{\em emergent}, robustly collective forms of memory. In general, while social 
context is seen to influence remembering, the act of remembering is held to 
be individual. However the literature on {\sf collective intentionality} (\cite{Tuo}) 
considers collective memory as a form of collective attention to the past. The 
concepts underlying the model we discuss below are greatly inspired by the 
{\em ``we-mode''}, discussed by the philosopher {\em Raimo Tuomela}, though we 
place an additional emphasis on {\sf patterns} of social interaction (rather
than the interactions themselves).

If we admit the existence of emergent collective memory, the major question
is whether the processes of social collective memory resemble in any way the 
processes of individual memory or that of small groups (like families) studied
extensively by psychologists. In an astonishing thesis from an interdisciplinary
collaboration of neurobiology, medicine, cognitive science and anthropolgy,
Anastasio at al (\cite{Anast}) assert that these two processes are in fact the
same. The model we present below, taking automata as memory representations,
attempts to build a `social automaton' (roughly speaking) in this spirit, showing
a correspondence of processes. 

The mechanism that we use in the model is closely related to that of {\em
majority dynamics} (\cite{KKT}) and {\em spread of influence} (\cite{MNT}) 
studied in the analysis of social networks. There are also influential logical
studies of diffusion in social networks (\cite{BCRS}). In other papers, we have studied
similar phenomena in the context of {\em large games} (\cite{PR1}, \cite{PR2}).
While there is a correspondence at an intuitive level between memory systems 
and large games, a formal correspondence would
lead to many potential applications.

The logic we propose here is a simple linear time temporal logic (with past)
with variables and bounded quantification. There is extensive logical
literature on the role of memory in multi-agent epistemic reasoning. For 
instance, the notion of bounded-recall plans has been studied in (\cite{DN},
\cite{BLY}, \cite{AL}), to mention some recent works. However, this 
literature is principally on the interactions between individual memories;
our point of departure is in studying emergent collective memory as an
entity in itself. Moreover, the logic we discuss here is extremely poor in
logical machinery, in comparison: there is no agency and no ascription
of epistemic attitudes. Rather, we discuss only the temporal evolution of
signalling interaction and the stablity of signals. On the other hand, logical
studies of social networks such as (\cite{BCRS}) and {\cite{LSG}) are
closer in spirit to the automaton model we present, though the logics
themselves are different. 

\section{A model}
Let $N$ denote a fixed finite set of agent names. Let ${\cal C} \subseteq
2^N$ be a nonempty set of nonempty subsets of $N$, referred to as 
{\em neighbourhoods} over $N$. We assume $|N| > 2$.

For presenting the model we will make some simplifying assumptions. We
fix a finite {\bf signal alphabet} $\Gamma$ common to all agents. Let
$\Gamma = \{\gamma_1, \ldots, \gamma_m\}$. 

Let $I \in {\cal C}$ and $|I| = k$. A {\bf distribution} over $I$ is an
$m$ tuple of integers $\pro =(y_1, \ldots, y_m)$ such that $y_j \geq 0$ 
and $\Sigma_{j=1}^{m}y_j = k,\ \ 1\leq j\leq m$. That is, the $j$th 
component of $\pro$ gives the number of agents in the neighbourhood $I$ 
who give signal $\gamma_j$. Let $\proset[I]$ denote the set of all signal
distributions of a neighbourhood $I$ and let $\proset = 
\bigcup_{I \in {\cal C}} \proset[I]$.

The main idea is this. All agents initially receive an external input and 
assume some state. At each state, an agent produces a signal. Interactions 
occur in neighbourhoods nondeterministically, and an agent who is a 
member of many, could be interacting in different neighbourhoods (though
every interaction at any instant is confined to one neighbourhood). Each 
interaction induces a state transition that is determined only by the 
distribution of signals: it does not depend on who is signalling
what, but how many are producing each signal. Such interactions keep
occurring repeatedly until a {\sf stable} configuration is reached.

Below, for $I \in {\cal C}$, we use the notation $\Gamma^I$ for a vector 
of signals, one signal for each of the agents in $I$. Note that every such vector
induces a distribution over $\Gamma$ in $\proset[I]$. 

We will consider systems of agents below where the set of possible
states $Q$ is uniform for all agents. Thus a state transition in $Q \times Q$
is also possible uniformly for all agents. By a triple $(\gamma, q, q')$ we
mean that an agent who is signalling $\gamma$ changes state from $q$ 
to $q'$. $\sigma: \Gamma \to (Q \times Q)$ specifies such a signal based
change of state. Let $\Sigma$ denote the collection of such maps.

\begin{definition}
A {\sf memory system} over ${\cal C}$ is a tuple $M = (Q, \delta, \iota, \omega)$,
where 
\begin{itemize}
\item $Q$ is a finite set of memory states,
\item $\iota: N \to Q$ is the initial state, 
\item $\omega: Q \to \Gamma$ is the {\em signalling function}, and 
\item $\delta$ is a finite family of {\em transition relations} $\delta_I \subseteq 
(\proset[I] \times \Sigma)$, where $I \in {\cal C}$.
\end{itemize} 
\end{definition}

\subsection{Dynamics}
A configuration $\chi$ is an element of $Q^N$. Let $\omega(\chi)$ denote the
vector in $\Gamma^N$ induced by $\omega$. For $I \in {\cal C}$ and let
$\pro_I$ be a distribution of signals induced by the vector $\omega(\chi)$
restricted to $I$. 

\begin{definition}
We say that an $I$-interaction is {\em enabled} at $\chi$ if there is a transition 
$(\pro, \sigma)$ in $\delta_I$ where $\pro$ is the distribution induced by
$\omega(\chi)$ and $\sigma \in \Sigma$ is a signal-based change of state.

The effect of the transition is determined  by $\sigma$ and the new configuration 
$\chi'$ is given by: 
\[\chi'(j) = q', {\rm ~where~} j \in I, \sigma(\omega(\chi(j))= (\chi(j), q') \]
and $\chi'(j) = \chi(j)$, otherwise. Thus, we have a transition $(\chi, I, \chi')$ on
configurations labelled by neighbourhoods.
\end{definition}

The dynamics of $M$ is then given by a {\sf configuration graph} $G_M$ 
whose vertices are configurations and edges are labelled by neighbourhoods: 
an edge $(\chi, I, \chi')$ is present if an $I$-interaction is enabled at $\chi$ by
a transition in $\delta$ with resulting configuration $\chi'$ as above.
Note that $\iota$ specifies an initial configuration $\chi_0$. A {\sf history}
$\rho$ is any finite or infinite path in $G_M$ starting from $\chi_0$. When
$\rho$ is finite, $|\rho|$ denotes its length. Let 
${\cal H}_M$ denote the set of all maximal histories of $M$. 

\subsection{Collective memory in $M$}
Consider a history $\rho = \chi_0 \chi_1 \ldots$ of system $M$. 
We say that signal $\gamma$ is {\sf eventually stable} for neighbourhood 
$I$ in $\rho$  if there exists $k$ such that for all $\ell \geq k$, (when $\rho$ 
is finite, for all $\ell$ such that $k \leq \ell \leq |\rho|$), and for all $j \in I$, 
$\omega(\chi_\ell(j)) = \gamma$. 

We say that $\gamma$ is in {\sf collective memory} in $\rho$ if it is 
eventually stable for $N$ in $\rho$: that is, no matter what interactions 
take place, all agents remain in states that emit signal $\gamma$. 

When we have stable configurations, we see them as formation of collective
memory. For this it is of course essential that a history allows for signalling
to spread across neighbourhoods; if some neighbourhoods never interact,
then signalling can remain confined within pockets. So we consider
{\sf spanning} histories where we impose the condition that every interaction
that is infinitely often enabled (according to the transition rule)  in an
infinite history takes place infinitely often. This is a typical {\sf fairness}
condition used in the theory of computation, but weaker, or different, conditions 
may be sufficient for many systems. For instance, we might ask that the
union of neighbourhoods in a history span all of $N$; this merely says
that all the agents have interacted at least once. Note that this depends
on $\delta$: the distributions specified may already disable some agents
or neighbourhoods from ever interacting. 

\begin{definition}
We say that the system $M$ {\em supports emergence of collective memory} 
if, for every {\em spanning} history  $\rho$ of $M$, there exists a signal 
$\gamma_\rho$ that is in collective memory in $\rho$.
\end{definition}

\subsection{An example}
Consider a system with two signals $\{g,b\}$, standing for
``good'' and ``bad''. There are only two states: $G$ and $B$, signalling
$g$ and $b$ respectively. Initially every agent perceives some global 
event as good or bad. Thus $\iota$ is an arbitrary distribution of
signals in the system.

The transition rule is simple: for any neighbourhood $I$, if more than 
half in $I$ signal $x$, then all agents in $I$ signal $x$ in  the new state.  
If they are exactly even, they continue evenly matched. Let $I$ be a
neighbourhood with $k$ agents.
$\delta_I = \{(m,n), (g, q,G), (b,q,G) | m > n, m+n=k\} \cup 
\{(m,n), (g, q,B), (b,q,B) | m < n, m+n=k\} \cup
\{(m,m), (s,q,q) | k = 2m, s \in \{g,b\}\}$.

Now we can see that whether either of the signals becomes stable in
a history depends on both the initial perception as well as which $I$-interactions
are enabled. If odd-sized neighbourhoods can interact, a signal will
begin to dominate. When the initial distribution is exactly even,
and the interacting neighbourhoods are always split evenly, neither
signal dominates the other. 

On the other hand, suppose that a large fraction of the population 
receives the signal $g$, but $\delta$ only enables neighbourhoods with the
majority signalling $b$ and a minority signalling $g$ to interact. Then the
signal $b$ emerges as a stable signal, despite the initial distribution.
This can be seen as the influence of social structures on collective
memory.

Such behavioural analysis is common in the study of {\em runaway
phenomena} (\cite{Ban}) and so-called {\em informational cascades}
(\cite{Zoe}).

\subsection{Some subclasses}
Consider a memory system in which all interactions are constrained to
be {\sf pairwise}. In such a system, the distribution profiles are entirely
irrelevant, since given a pair of agents, there are four possible pairs
of signals, determined by their states, and we only need to specify
the resulting pair of states. Thus $\delta \subseteq Q^4$. Such systems
have been studied as {\bf population protocols} (\cite{Pop}), for which
a number of technical results are known.

We also have other interesting subclasses of systems: for instance,
those where the transition relation is presented as $\delta_k$ where
$1 < k \leq N$. That is, only the size of a neighbourhood determines
whether it can interact and not the identities of agents in it. (Note that
even this restriction does not rule out the predatory dominance of
a signal, as illustrated in the example above.)

Of great interest to social systems is where ${\cal C}$ represents a
{\sf hierarchy}: we have an ordering $<$ on $N$ and impose the
condition that  all $I \in {\cal C}$ are downward-closed with respect
to the ordering relation. 

However, note that these subclasses only constrain interaction 
structure and not the memory updates. This may be consonant
with the discussion on collective memory in social theory, whereby
social structures (and belief systems) constrain opportunity and
influence, but what persists in social memory may well be
impervious to social structures. 

\subsection{Computational power}
Note that when $|\Gamma| = d$, every distribution over $\Gamma$ is
a $d$-dimensional vector in $\nat^d$. The initial state $\iota$ specifies
such a distribution. Now consider a spanning history that is stably 
signalling $\gamma$; we can consider this the output of the memory
system for that history. Viewed thus, every system that supports
emergence of collective memory can be said to {\em compute}
a function from $\proset \to \Gamma$. Alternatively, we can consider
such a function to be a predicate over $\nat^d \times \{1,\dots,d\}$.
Thus we can speak of arithmetical {predicates computable by memory 
systems}.

An important theorem in the study of population protocols guides us
to the study of what memory systems can compute. In the study of
population protocols, systems come with an {\em output function},
mapping to the two element output alphabet $\{0,1\}$ (without 
loss of generality). In this case we can consider the population 
protocol to be computing a predicate over $\nat^d$. 

Recall that a {\em semi-linear set} is a subset of $\nat^d$ that is a
finite union of {\em linear} sets of the form 
$\{{\bf b} + k_1 {\bf a_1} + \ldots + k_m {\bf a_m} \mid k_1, \ldots, k_m \in \nat\}$,
where ${\bf b} \in \nat^d$ and ${\bf a_1}  \ldots {\bf a_m}$ are $d$-dimensional
basis vectors. 

\begin{theorem}
{\bf \cite{Power}}: A predicate is computable by a population protocol
iff it is semi-linear.
\end{theorem}

An alternative characterization of these predicates is that they
can be expressed in first-order Presburger arithmetic, which is first order 
arithmetic on the natural numbers with addition but not multiplication.

\begin{theorem}
Given a memory system $M$, checking whether $M$ supports emergence of 
collective memory is decidable.
Moreover the class of predicates computable by memory systems is
exactly that of population protocols.
\end{theorem}

There are two parts to the proof. We construct a {\sf Parikh automaton} 
(\cite{Kla-R} that represents the configuration space of the memory system
and reduce the check for stability of signals in the system to the nonemptiness
problem for the associated Parikh automaton. This gives us the required
decision procedure. 

In the process we show that every predicate computed by a memory
system is semi-linear. By the earlier theorem, such a predicate can
be computed by a population protocol. Conversely, since population
protocols are a subclass of memory systems, the predicates computed
by the former are computable by memory systems as well. (There are
some details related to the output function, and the restriction to
spanning histories, which complicate the construction a little.)

To get an intuitive idea of the construction, we define Parikh automaton
below, which in turn needs the definition of Presburger arithmetic.

Firstly, let $\Delta = \{a_1, \ldots a_m\}$ be any finite alphabet, and 
$w \in \Delta^*$. The Parikh image of $w$ counts the number of 
occurrences of each letter of the alphabet in $w$.  Formally, we
have the map $\pi: \Delta^* \to \nat^m$ given by: $\pi(a_i) = e_i$
and $\pi(uv) = \pi(u) + \pi(v)$, where $e_i$ is the unit vector of length
$m$ where the $i^{th}$ coordinate is $1$. 

Clearly the map $\pi$ can be lifted to alphabets of the form 
$\Delta \times D$ where $D \subseteq \nat^d$: $\pi(a_i, \hat{k}) = \hat{k}$
and  $\pi(uv) = \pi(u) + \pi(v)$. We can also consider the alphabetic
projection into $D^*$: $\lambda(a_i, \hat{k}) = a_i$ and 
$\lambda(uv) = \lambda(u) \lambda(v)$.

Presburger arithmetic is first order logic with the only atomic formulas
of the form $t~rel~t'$ where $t$ and $t'$ are terms, and $rel \in \{>, <, \geq, \leq\}$.
Terms are built from two constants $0$ and $1$, and variables, using
addition and $n \cdot t$ where $n \in \nat$. Formulas are interpreted
over the structure ${\cal N} = (\nat, +, \cdot, 0, 1)$. When $\phi(\hat{x})$
is a formula with free variables $\hat{x}$ the notion ${\cal N}, \hat{k} \models
\phi(\hat{x})$ is defined in the standard fashion.

Given $D \subseteq \nat^d$ and $L \subseteq (\Delta \times D)^*$, and
a formula $\phi$ of Presburger arithmetic, we define $L \lceil_\phi
= \{\lambda(w) \mid {\cal N}, \pi(w) \models \phi(\hat{x})\}$.

\begin{definition}
A {\bf Parikh automaton} of dimension $d > 0$ is a pair $(A, \phi)$ where
$\phi(x_1, \ldots, x_d)$ is a formula of Presburger arithmetic over $d$
variables, and $A$ is a finite word automaton with the finite alphabet 
$\Delta \times D$ where $D \subseteq \nat^d$. We say that  $(A, \phi)$
recognizes $L(A, \phi) = L(A) \lceil_\phi$, where $L(A)$ is the language
recognized by the automaton $A$.
\end{definition}

For the construction we need, there are some points to note. Configurations
of memory systems carry states, from which we compute signal distributions
which cause state changes. For the Parikh automaton, transitions are
labelled by distributions. More importantly we need to carry the neighbourhood
based signalling in the transitions of the Parikh automaton. While these are
matters of detail, the harder part of the proof is the definition of the formula
of Presburger arithmetic, for which we closely follow the proof method for population
protocols (\cite{Power}).

\section{A logic for the rationale}
We now turn our attention to our principal logical interest, namely, the
rationale by which agents decide what signals are chosen. This is surely
complex, depending on the systems being modelled. Here we propose
a minimal logic, in which agents evaluate signals based on their own
evaluation of the current state and their evaluation of signals from the
neighbourhood, depending on the signal distributions. 

Let $V$ be a countable set of variables. Let the terms of
the logic be defined as
$$\term ::= i\ |\ x,\ i\in N,\ x\in V$$
That is, a term is either an agent name or a variable (which takes
agents as its values). Let $\prop$ denote a countable set of
atomic propositional symbols.

The formulas of the logic are built using the following syntax:

\[
\begin{split}
\Phi ::= & \tau_1=\tau_2\ |\ \tau \in I \ |\  p@\tau, p\in \prop\ |\
\gamma@\tau, \gamma \in \Gamma\ |\
\neg\varphi\ |\\
& \varphi_1\lor\varphi_2\ |\ \ominus\varphi\ |\ \bigcirc\varphi\ |\
\diamondmin\varphi\ |\ \Diamond\varphi\ |\ \num x\cdot\varphi(x)~op~k
\end{split}
\]
where $\tau_1$ and $\tau_2$ are terms, $I \in {\cal C}$, $op \in \{=, \neq, <, \leq, >, \geq\}$ and $k \in \{0, \ldots, |N|\}$.

The formula $\tau \in I $ asserts that the agent denoted by the term $\tau$ is
a member of the neigbourhood $I$. $p@\tau$ asserts that the condition $p$
holds for agent $\tau$, and $\gamma@\tau$ specifies that $\tau$ is
signalling $\gamma$ at that instant. $\num$ is a counting
quantifier, and $\num x\cdot\varphi(x) \geq k$ (for example) says that
at least $k$ agents support the assertion $\phi$ at the instant. The modalities
$\ominus$ and $\bigcirc$ denote the predecessor and successor instants,
whereas $\diamondmin$ and $\Diamond$ denote some time in the past
and some time in the future, respectively. Their dual modalities are denoted
$\boxmin$ and $\Box$ respectively. We talk of free and bound occurrences 
of variable $x$ in formula $\varphi$ in the standard manner. $\varphi$ is said 
to be a sentence if it has no free occurrences of variables.

The existential and universal quantifiers are defined easily: $\exists x \cdot \phi(x) 
= \num x\cdot\varphi(x) > 0$ and $\forall x  \cdot \phi(x)  = \neg \exists x \neg \phi(x)$.
We use the abbreviation $\gamma@I = \forall x \cdot (x \in I \supset \gamma@x)$
to denote that all agents in the neighbourhood signal $\gamma$. 

The formula $\Box \forall x \cdot \gamma@x$ is denoted $stable(\gamma)$. The
formula $\orover_{\gamma \in \Gamma} stable(\gamma)$ is special and is called
{\bf emergence}. 

The semantics is defined on histories. A model is a tuple $(M, \val, \varmap)$, 
where $M$ is a memory system, $\varmap: V \to N$ denote an assignment of agent variables to agent names and
$\val: Q \to 2^{\prop}$ is the propositional valuation map. $\val$ is lifted to configurations
by the map $\hat{\val}: Q^N \to (N \to 2^{\prop})$  by: $\hat{\val}(\chi) (i)
= \val(\chi(i))$. 

Let $\rho \in {\cal H}_M$ be a finite or infinite history $\rho = \chi_0 \chi_1 \ldots$. 
The notion that $\rho, k \models \phi$ is defined in the standard fashion, for
$k \geq 0$. 

\begin{itemize}
\item $\rho, k \models \tau_1=\tau_2$ iff $\varmap(\tau_1) = \varmap(\tau_2)$.
\item $\rho, k \models \tau \in I$ iff $\varmap(\tau) \in I$.
\item $\rho, k \models p@\tau$ iff $p \in \hat{val}(\rho_k)(\varmap(\tau))$.
\item $\rho, k \models \gamma@\tau$ iff $\omega(\rho_k(\tau)) = \gamma$.
\item $\rho, k \models  \neg \varphi$ iff  $\rho, k \not\models \varphi$.
\item $\rho, k \models  \varphi_1\lor \varphi_2$ iff $\rho, k \models  \varphi_1$ or
$\rho, k \models  \varphi_2$.
\item $\rho, k \models \ominus\varphi$ iff $k > 0$ and $\rho, k-1 \models \varphi$.
\item $\rho, k \models \bigcirc\varphi$ iff there exists a successor instant in the
history and $\rho, k+1 \models \varphi$.
\item $\rho, k \models \diamondmin\varphi$ iff there exists $\ell \leq k$ such that
$\rho, \ell \models \varphi$.
\item $\rho, k \models \Diamond \varphi$ iff there exists $\ell \geq k$ such that
$\rho, \ell \models \varphi$.
\item $\rho, k \models \num x\cdot \varphi(x)~op~k$ iff $\mid \{j \mid \rho, k \models
\varphi[j/x]\}\mid ~op~k$.
\end{itemize}

The notions of satisfiability and validity are standard. Given a model $(M, \val)$ and
a sentence $\phi$ by $M \models \varphi$ we denote that for all histories $\rho$ in
${\cal H}_M$, $\rho, 0 \models \varphi$. 

\subsection{Examples}
It is easily seen that specific distributions can be described in the logic.
For instance consider the distribution where there are 100 agents, 30
of whom signal  $a$ and the rest signal $b$. We can specify this as:
$$ \num x \cdot \gamma_a@x = 30 \land \num x \cdot \gamma_b@x = 70 $$
With inequalities, classes of distributions that lead to the same signalling'
behaviour can be specified. 

Further the structure of interaction in histories can be constrained in the logic:
$$\exists x \cdot (x \in I \supset \Box (x \in I))$$
This asserts that once an agent participates in an interaction in the history,
it continues to participate in every interaction in the history. (In social theory,
such persistent actors in the structure are associated with memorials that
keep reminding everyone of a memory token.)

The state transition structure of memory systems can be described only to
a limited extent since the logic is first order and the state information which
may include modular counting of signals cannot be expressed in it. However,
propositional updates based on signal distributions can be specified.

The logic can describe various kinds of signalling schemes by agents.

\begin{itemize}
\item Signal $a$ and $b$ alternatively:
$$\Box [(\ominus \gamma_a@i \supset \gamma_b@i) \land 
(\ominus \gamma_b@i \supset \gamma_a@i)]$$

\item If more than 5 agents in my neighbourhood previously signalled $a$
then I signal $a$:
$$(i \in I \land \num x. (x \in I \land \ominus \gamma_a@x)  > 5) \supset \gamma_a@i$$

\item $\gamma$ is collective memory:
$$\Diamond \Box (\forall x. \gamma@x)$$

\item $\gamma$ is collective amnesia:
$$\Diamond \Box (\forall x. \neg \gamma@x)$$

\end{itemize}

However, in terms of validities, the logic has little structure to force validities
beyond that of linear time temporal logic. The interest of the logic is mainly
in its role as a specification language for requirements on memory systems,
and hence we are more interested in checking whether a specific memory
system satisfies such a specification.

\subsection{Model checking}
In general, first order temporal logics are highly undecidable and non-axiomatizable.
However, what we have here is bounded quantification, since the set of agents 
over which we quantify, is fixed and finite. So we can effectively eliminate
quantifiers and translate formulas into propositional linear time temporal logic.
Thus satisfiability is decidable via automaton construction, but yet, extracting 
a memory system from the formula automaton has some interesting details.

The model checking problem for the logic asks, given a model $(M, \val)$ and
a sentence $\varphi$, whether $M \models \varphi$. 

\begin{theorem}
The model checking problem for the logic is decidable in time linear in $M$ and
singly exponential in $\varphi$. 
\end{theorem}

In this case, we construct a Parikh automaton that represents the
configuration space of the memory system, and take its product with
the formula automaton associated with the given sentence $\varphi$.
The construction is straightforward, though not entirely trivial.

\section{Discussion}
We began with the intention of {\sf reasoning} about collective memory.
How do systems of signalling in neighbourhoods embody reasoning?

Firstly, it should be clear that the history model and the logical
language are rich enough to talk about the ``remembrance of 
things past''. However, the model assumes perfect observability
for agents within neighbourhoods within a fixed interaction
structure. Both of these assumptions need to be relaxed, 
leading to epistemic logics.

Such a logical exercise is not sufficient in itself to uncover the 
process of {\em stipulation} mentioned by Sontag, or the {\em interdependence}
between memory and reasoning demanded by Wittgenstein. However, in our
opinion, the model holds considerable promise. For achieving the richness
required, we hold two features to be essential: reinforcement of memory 
that comes through repeated interactions inside local neighbourhoods,
but not confined to those neighbourhoods; complex social rules that
determine influence in signalling. Elements of both are present in this model.

Further, while we have presented simple stability as the basic notion of
collective memory which is persistent, we can symmterically study notions
like {\em collective amnesia} in the model whereby signals predominant
in the system lose out in interactions and ultimately vanish. Incorporating
both, to study collective memory of some events while at the same time
forgetting others presents no logical difficulties but makes the model 
considerably complex to reason about, requiring a structural insight
for simplification.

Moreover, we have taken collective memory to mean the entire system.
We can instead parameterize such memory by a subset of agents to get 
group notions of memory; this is relavant in the study of notions such
as common ground for communication (\cite{Clark}).

It is interesting to consider dynamic memory systems, whereby endogenous 
changes in signalling behaviour can lead to altering interaction structures 
leading to new update rules. In particular, neighbourhoods need not be static,
but may expand and contract. This is analogous to dynamic form games
as we have studied elsewhere (\cite{PR1}).

Social choice theory offers a variety of methods for aggregation of information
from individuals for collectives, but these are static rules. Automata models 
such as the ones studied here can offer dynamic methods of aggregation
whereby we formulate `local' aggregation rules which are applied repeatedly.
Whether this can lead to meaningful insight for social theories remains
to be seen, offering interesting technical questions for study in the meanwhile.

Interactions in the model are nondeterministic. Stochastic models of interactions
may be more appropriate for social behaviour, relevant to social network
studies such as (\cite{KKT}, \cite{MNT}). However, logicising the rationale
of such interaction would perhaps need a different approach.

\section{Acknowledgement}
We thank Oliver Roy and Wang Yi for discussions on the theme, and the TARK
reviewers for thoughtful remarks and helpful comments.


\nocite{*}
\bibliographystyle{eptcs}
\bibliography{ref}

\begin{thebibliography}{10}
\providecommand{\bibitemdeclare}[2]{}
\providecommand{\surnamestart}{}
\providecommand{\surnameend}{}
\providecommand{\urlprefix}{Available at }
\providecommand{\url}[1]{\texttt{#1}}
\providecommand{\href}[2]{\texttt{#2}}
\providecommand{\urlalt}[2]{\href{#1}{#2}}
\providecommand{\doi}[1]{doi:\urlalt{http://dx.doi.org/#1}{#1}}
\providecommand{\bibinfo}[2]{#2}

\bibitemdeclare{article}{AL}
\bibitem{AL}
\bibinfo{author}{Natasha \surnamestart Alechina\surnameend} \&
  \bibinfo{author}{Brian \surnamestart Logan\surnameend}
  (\bibinfo{year}{2010}): \emph{\bibinfo{title}{Belief ascription under bounded
  resources}}.
\newblock {\sl \bibinfo{journal}{Synth.}}
  \bibinfo{volume}{173}(\bibinfo{number}{2}), pp. \bibinfo{pages}{179--197},
  \doi{10.1007/s11229-009-9706-6}.

\bibitemdeclare{book}{Anast}
\bibitem{Anast}
\bibinfo{author}{Thomas~J \surnamestart Anastasio\surnameend},
  \bibinfo{author}{Kristen~Ann \surnamestart Ehrenberger\surnameend},
  \bibinfo{author}{Patrick \surnamestart Watson\surnameend} \&
  \bibinfo{author}{Wenyi \surnamestart Zhang\surnameend}
  (\bibinfo{year}{2012}): \emph{\bibinfo{title}{Individual and collective
  memory consolidation: Analogous processes on different levels}}.
\newblock \bibinfo{publisher}{MIT Press}, \doi{10.7551/mitpress/9173.001.0001}.

\bibitemdeclare{article}{Power}
\bibitem{Power}
\bibinfo{author}{Dana \surnamestart Angluin\surnameend}, \bibinfo{author}{James
  \surnamestart Aspnes\surnameend}, \bibinfo{author}{David \surnamestart
  Eisenstat\surnameend} \& \bibinfo{author}{Erik \surnamestart
  Ruppert\surnameend} (\bibinfo{year}{2007}): \emph{\bibinfo{title}{The
  computational power of population protocols}}.
\newblock {\sl \bibinfo{journal}{Distributed Computing}}
  \bibinfo{volume}{20}(\bibinfo{number}{4}), pp. \bibinfo{pages}{279--304},
  \doi{10.1007/s00446-007-0040-2}.

\bibitemdeclare{inbook}{Pop}
\bibitem{Pop}
\bibinfo{author}{James \surnamestart Aspnes\surnameend} \&
  \bibinfo{author}{Eric \surnamestart Ruppert\surnameend}
  (\bibinfo{year}{2009}): \emph{\bibinfo{title}{An Introduction to Population
  Protocols}}, pp. \bibinfo{pages}{97--120}.
\newblock \bibinfo{publisher}{Springer Berlin Heidelberg},
  \bibinfo{address}{Berlin, Heidelberg}, \doi{10.1007/978-3-540-89707-1_5}.

\bibitemdeclare{article}{BCRS}
\bibitem{BCRS}
\bibinfo{author}{Alexandru \surnamestart Baltag\surnameend},
  \bibinfo{author}{Zo{\'{e}} \surnamestart Christoff\surnameend},
  \bibinfo{author}{Rasmus~K. \surnamestart Rendsvig\surnameend} \&
  \bibinfo{author}{Sonja \surnamestart Smets\surnameend}
  (\bibinfo{year}{2019}): \emph{\bibinfo{title}{Dynamic Epistemic Logics of
  Diffusion and Prediction in Social Networks}}.
\newblock {\sl \bibinfo{journal}{Studia Logica}}
  \bibinfo{volume}{107}(\bibinfo{number}{3}), pp. \bibinfo{pages}{489--531},
  \doi{10.1007/s11225-018-9804-x}.

\bibitemdeclare{article}{Ban}
\bibitem{Ban}
\bibinfo{author}{Abhijit~V. \surnamestart Banerjee\surnameend}
  (\bibinfo{year}{1992}): \emph{\bibinfo{title}{A Simple Model of Herd
  Behaviour}}.
\newblock {\sl \bibinfo{journal}{The Quarterly Journal of Economics}}
  \bibinfo{volume}{107}(\bibinfo{number}{3}), pp. \bibinfo{pages}{797--817},
  \doi{10.2307/2118364}.

\bibitemdeclare{inbook}{Barash}
\bibitem{Barash}
\bibinfo{author}{Jeffrey~Andrew \surnamestart Barash\surnameend}
  (\bibinfo{year}{2016}): \bibinfo{publisher}{University of Chicago Press},
  \doi{10.7208/9780226399294-toc}.

\bibitemdeclare{inproceedings}{BLY}
\bibitem{BLY}
\bibinfo{author}{Francesco \surnamestart Belardinelli\surnameend},
  \bibinfo{author}{Alessio \surnamestart Lomuscio\surnameend} \&
  \bibinfo{author}{Emily \surnamestart Yu\surnameend} (\bibinfo{year}{2020}):
  \emph{\bibinfo{title}{Model Checking Temporal Epistemic Logic under Bounded
  Recall}}.
\newblock In: {\sl \bibinfo{booktitle}{Proceedings AAAI 2020}},
  \bibinfo{publisher}{{AAAI} Press}, pp. \bibinfo{pages}{7071--7078},
  \doi{10.1609/aaai.v34i05.6193}.

\bibitemdeclare{article}{Zoe}
\bibitem{Zoe}
\bibinfo{author}{Zo{\'{e}} \surnamestart Christoff\surnameend} \&
  \bibinfo{author}{Jens~Ulrik \surnamestart Hansen\surnameend}
  (\bibinfo{year}{2015}): \emph{\bibinfo{title}{A logic for diffusion in social
  networks}}.
\newblock {\sl \bibinfo{journal}{J. Appl. Log.}}
  \bibinfo{volume}{13}(\bibinfo{number}{1}), pp. \bibinfo{pages}{48--77},
  \doi{10.1016/j.jal.2014.11.011}.

\bibitemdeclare{inbook}{Clark}
\bibitem{Clark}
\bibinfo{author}{Herbert~H \surnamestart Clark\surnameend} \&
  \bibinfo{author}{Susan~E \surnamestart Brennan\surnameend}
  (\bibinfo{year}{1991}): \emph{\bibinfo{title}{Grounding in communication}}.
\newblock \bibinfo{publisher}{American Psychology Association},
  \doi{10.1037/10096-006}.

\bibitemdeclare{inproceedings}{DP}
\bibitem{DP}
\bibinfo{author}{C.~\surnamestart Daskalakis\surnameend} \&
  \bibinfo{author}{C.~H. \surnamestart Papadimitriou\surnameend}
  (\bibinfo{year}{2007}): \emph{\bibinfo{title}{Computing equilibria in
  anonymous games}}.
\newblock In: {\sl \bibinfo{booktitle}{Proceedings of the 48th symposium on
  Foundations of Computer Science (FOCS)}}, \bibinfo{publisher}{IEEE Computer
  Society Press}, pp. \bibinfo{pages}{83--93}, \doi{10.1109/FOCS.2007.24}.

\bibitemdeclare{article}{DN}
\bibitem{DN}
\bibinfo{author}{Kaya \surnamestart Deuser\surnameend} \&
  \bibinfo{author}{Pavel \surnamestart Naumov\surnameend}
  (\bibinfo{year}{2020}): \emph{\bibinfo{title}{On composition of
  bounded-recall plans}}.
\newblock {\sl \bibinfo{journal}{Artif. Intell.}} \bibinfo{volume}{289}, p.
  \bibinfo{pages}{103399}, \doi{10.1016/j.artint.2020.103399}.

\bibitemdeclare{article}{Gilb}
\bibitem{Gilb}
\bibinfo{author}{Margaret \surnamestart Gilbert\surnameend}
  (\bibinfo{year}{1987}): \emph{\bibinfo{title}{Modelling Collective Belief}}.
\newblock {\sl \bibinfo{journal}{Synthese}}
  \bibinfo{volume}{73}(\bibinfo{number}{1}), pp. \bibinfo{pages}{185--204},
  \doi{10.1007/BF00485446}.

\bibitemdeclare{book}{Halb}
\bibitem{Halb}
\bibinfo{author}{Maurice \surnamestart Halbwachs\surnameend}
  (\bibinfo{year}{1950}): \emph{\bibinfo{title}{The Collective Memory}}.
\newblock \bibinfo{publisher}{Harper and Row}, \bibinfo{address}{New York}.

\bibitemdeclare{article}{KKT}
\bibitem{KKT}
\bibinfo{author}{David \surnamestart Kempe\surnameend}, \bibinfo{author}{Jon
  \surnamestart Kleinberg\surnameend} \& \bibinfo{author}{\'{E}va \surnamestart
  Tardos\surnameend} (\bibinfo{year}{2015}): \emph{\bibinfo{title}{Maximizing
  the Spread of Influence through a Social Network}}.
\newblock {\sl \bibinfo{journal}{Theory of Computing}}
  \bibinfo{volume}{11}(\bibinfo{number}{4}), pp. \bibinfo{pages}{105--147},
  \doi{10.4086/toc.2015.v011a004}.

\bibitemdeclare{inproceedings}{Kla-R}
\bibitem{Kla-R}
\bibinfo{author}{Felix \surnamestart Klaedtke\surnameend} \&
  \bibinfo{author}{Harald \surnamestart Rue{\ss}\surnameend}
  (\bibinfo{year}{2003}): \emph{\bibinfo{title}{Monadic Second-Order Logics
  with Cardinalities}}.
\newblock In \bibinfo{editor}{Jos C.~M. \surnamestart Baeten\surnameend},
  \bibinfo{editor}{Jan~Karel \surnamestart Lenstra\surnameend},
  \bibinfo{editor}{Joachim \surnamestart Parrow\surnameend} \&
  \bibinfo{editor}{Gerhard~J. \surnamestart Woeginger\surnameend}, editors:
  {\sl \bibinfo{booktitle}{Automata, Languages and Programming}},
  \bibinfo{publisher}{Springer Berlin Heidelberg}, pp.
  \bibinfo{pages}{681--696}, \doi{10.1007/3-540-45061-0_54}.

\bibitemdeclare{article}{LSG}
\bibitem{LSG}
\bibinfo{author}{Fenrong \surnamestart Liu\surnameend}, \bibinfo{author}{Jeremy
  \surnamestart Seligman\surnameend} \& \bibinfo{author}{Patrick \surnamestart
  Girard\surnameend} (\bibinfo{year}{2014}): \emph{\bibinfo{title}{Logical
  dynamics of belief change in the community}}.
\newblock {\sl \bibinfo{journal}{Synthese}}
  \bibinfo{volume}{191}(\bibinfo{number}{11}), pp. \bibinfo{pages}{2403--2431},
  \doi{10.1007/s11229-014-0432-3}.

\bibitemdeclare{article}{MNT}
\bibitem{MNT}
\bibinfo{author}{Elchanan \surnamestart Mossel\surnameend},
  \bibinfo{author}{Joe \surnamestart Neeman\surnameend} \&
  \bibinfo{author}{Omer \surnamestart Tamuz\surnameend} (\bibinfo{year}{2013}):
  \emph{\bibinfo{title}{Majority dynamics and aggregation of information in
  social networks}}.
\newblock {\sl \bibinfo{journal}{Autonomous Agents and Multi-Agent Systems}}
  \bibinfo{volume}{28}(\bibinfo{number}{3}), p. \bibinfo{pages}{408–429},
  \doi{10.1007/s10458-013-9230-4}.

\bibitemdeclare{inbook}{Pa}
\bibitem{Pa}
\bibinfo{author}{Rohit \surnamestart Parikh\surnameend} (\bibinfo{year}{2012}):
  \emph{\bibinfo{title}{What Is Social Software?}}, pp. \bibinfo{pages}{3--13}.
\newblock {\sl \bibinfo{series}{LNCS}} \bibinfo{volume}{7010},
  \bibinfo{publisher}{Springer}, \doi{10.4204/EPTCS}.

\bibitemdeclare{article}{PR1}
\bibitem{PR1}
\bibinfo{author}{Soumya \surnamestart Paul\surnameend} \&
  \bibinfo{author}{R.~\surnamestart Ramanujam\surnameend}
  (\bibinfo{year}{2013}): \emph{\bibinfo{title}{Dynamics of Choice restriction
  in Large Games}}.
\newblock {\sl \bibinfo{journal}{{IGTR}}}
  \bibinfo{volume}{15}(\bibinfo{number}{4}), \doi{10.1142/S0219198913400318}.

\bibitemdeclare{article}{PR2}
\bibitem{PR2}
\bibinfo{author}{Soumya \surnamestart Paul\surnameend} \&
  \bibinfo{author}{R.~\surnamestart Ramanujam\surnameend}
  (\bibinfo{year}{2014}): \emph{\bibinfo{title}{Subgames within Large Games and
  the Heuristic of Imitation}}.
\newblock {\sl \bibinfo{journal}{Stud Logica}}
  \bibinfo{volume}{102}(\bibinfo{number}{2}), pp. \bibinfo{pages}{361--388},
  \doi{10.1007/s11225-014-9549-0}.

\bibitemdeclare{article}{RP-G}
\bibitem{RP-G}
\bibinfo{author}{William~G. \surnamestart Roy\surnameend} \&
  \bibinfo{author}{Rachel \surnamestart Parker-Gwin\surnameend}
  (\bibinfo{year}{1999}): \emph{\bibinfo{title}{How Many Logics of Collective
  Action?}}
\newblock {\sl \bibinfo{journal}{Theory and Society}}
  \bibinfo{volume}{28}(\bibinfo{number}{2}), pp. \bibinfo{pages}{203--237},
  \doi{10.1023/A:1006946310119}.

\bibitemdeclare{article}{Rusu}
\bibitem{Rusu}
\bibinfo{author}{Mihai \surnamestart Rusu\surnameend} (\bibinfo{year}{2017}):
  \emph{\bibinfo{title}{Transitional Politics of Memory: Political Strategies
  of Managing the Past in Post-communist Romania}}.
\newblock {\sl \bibinfo{journal}{Europe-Asia Studies}}, pp.
  \bibinfo{pages}{260--282}, \doi{10.1080/09668136.2017.1380783}.

\bibitemdeclare{book}{Son}
\bibitem{Son}
\bibinfo{author}{Susan \surnamestart Sontag\surnameend} (\bibinfo{year}{2003}):
  \emph{\bibinfo{title}{Regarding the pain of others}}.
\newblock \bibinfo{publisher}{Picador}, \doi{10.3917/dio.201.0127}.

\bibitemdeclare{inbook}{Sutt}
\bibitem{Sutt}
\bibinfo{author}{John \surnamestart Sutton\surnameend} (\bibinfo{year}{2014}):
  \emph{\bibinfo{title}{Remembering as Public Practice: Wittgenstein, memory,
  and distributed cognitive ecologies}}.
\newblock \doi{10.1515/9783110378795.409}.

\bibitemdeclare{article}{Tamm}
\bibitem{Tamm}
\bibinfo{author}{Allard \surnamestart Tamminga\surnameend},
  \bibinfo{author}{Hein \surnamestart Duijf\surnameend} \&
  \bibinfo{author}{Frederik Van~De \surnamestart Putte\surnameend}
  (\bibinfo{year}{2020}): \emph{\bibinfo{title}{Expressivity results for
  deontic logics of collective agency}}.
\newblock {\sl \bibinfo{journal}{Synthese}}, pp. \bibinfo{pages}{260--282},
  \doi{10.1007/s11229-020-02597-0}.
\newblock \bibinfo{note}{Forthcoming}.

\bibitemdeclare{book}{Tuo}
\bibitem{Tuo}
\bibinfo{author}{Raimo \surnamestart Tuomela\surnameend}
  (\bibinfo{year}{2013}): \emph{\bibinfo{title}{Social Ontology: Collective
  Intentionality and Group Agents}}.
\newblock \bibinfo{publisher}{OUP USA}, \doi{10.1515/jso-2014-0040}.

\bibitemdeclare{book}{Witt}
\bibitem{Witt}
\bibinfo{author}{Ludwig \surnamestart Wittgenstein\surnameend}
  (\bibinfo{year}{1967}): \emph{\bibinfo{title}{Zettel}}.
\newblock \bibinfo{publisher}{Univ of California Press}.

\end{thebibliography}
\end{document}